\documentclass[aps,12pt,final,showpacs,showkeys,onecolumn,superscriptaddress]{revtex4}%
\usepackage{amsfonts}
\usepackage{amsmath}
\usepackage{amssymb}
\usepackage{graphicx}%
\providecommand{\U}[1]{\protect\rule{.1in}{.1in}}
\begin{document}

\title{Large-amplitude electron-acoustic solitons in a dusty plasma \\with kappa-distributed electrons}

\author{N. S. Saini}{
\affiliation{Department of Physics, Guru Nanak Dev University,
Amritsar-143005,
  India}
\author{A. Danehkar}{
\affiliation{Department of Physics and Astronomy, Macquarie
University, Sydney, NSW 2109,
  Australia}
\author{M. A. Hellberg}{
\affiliation{School of Physics, University of KwaZulu-Natal, Durban
4000, South Africa}
\author{I. Kourakis}{
\affiliation{Centre for Plasma Physics, Queen's University Belfast,
BT7 1 NN, Northern Ireland, UK}


\begin{abstract}
The Sagdeev pseudopotential method is used to investigate the
occurrence and the dynamics of fully nonlinear electrostatic
solitary structures in a plasma containing suprathermal hot
electrons, in the presence of massive charged dust particles in the
background. The soliton existence domain is delineated, and its
parametric dependence on different physical parameters is clarified.
\end{abstract}

\maketitle

%%%%%%%%%%%%%%%%%%%%%%%%%%%%%%%%%%%%%%%%%%%%
%% MAINMATTER
%%%%%%%%%%%%%%%%%%%%%%%%%%%%%%%%%%%%%%%%%%%%

%\section{<A section>}

%
% Abstract text.
%
Electron-acoustic (EA) waves occur in plasma which are characterized by a co-existence of
two distinct type of electrons: cool and hot electrons. A large
number of investigations have focused on electron acoustic solitary
waves in such plasma configuration \cite{dub,sv}. Abundant
observations suggest the presence of an excess population of
suprathermal electrons/ions in astrophysical environments, which is
efficiently modeled by a kappa-type distribution \cite{vas}.
Recently, these investigations have been extended to study EA excitations
in the presence of electrons modeled by a $\kappa$-distribution
\cite{sharmin,bal,ad}. Furthermore, the presence of dust particles
in space plasmas generates new modes, which modify the
characteristics of ion-acoustic \cite{shukla} and EA
\cite{yu,mam1} solitary waves.

We consider a four-component plasma, namely consisting of cold electrons,
suprathermal hot electrons, stationary ions and charged dust
particles. The cold electron fluid is described by the following normalized
equations \cite{ad}
\begin{eqnarray}
\frac{\partial n}{\partial t}&+&\frac{\partial (n u)}{\partial x} = 0,
\quad \frac{\partial u}{\partial t}+ u \frac{\partial u} {\partial x}=
\frac{\partial \phi}{\partial x} ,  \\
 \frac{\partial^{2}\phi}{\partial x^{2}}&=&-\left(  \eta+s\delta\right)
+n+\left(  \eta+s\delta-1\right)  \left[
1-\frac{\phi}{\kappa-3/2}\right] ^{-\kappa+1/2}, \label{e1}
\end{eqnarray}
where $n$ and $u$ denote the density and velocity of the cool
electron fluid normalized with respect to $n_{c,0}$ and the hot
electron thermal speed $c_{th}=\left(
k_{B}T_{h}/m_{e}\right) ^{1/2}$, respectively. The wave potential $\phi $ is scaled by $%
k_{B}T_{h}/e$, time and space by the plasma period $\omega
_{pc}^{-1}=\left( n_{c,0}e^{2}/\varepsilon _{0}m_{e}\right) ^{-1/2}$
and the characteristic length $\lambda _{0}=\left( \varepsilon
_{0}k_{B}T_{h}/n_{c,0}e^{2}\right)^{1/2}$, respectively. We define
the hot-to-cold electron charge density ratio $\gamma
=n_{h,0}/n_{c,0}$, the ion-to-cold electron charge density ratio
$\eta =Z_{i}n_{i,0}/n_{c,0}$, and the dust-to-cold electron charge
density ratio $\delta =Z_{d}n_{d,0}/n_{c,0}$. Here, suprathermality
is denoted by the spectral index $\kappa$, and $s=\pm1$ for
positive/negative dust. Charge neutrality at equilibrium yields
$\eta+s\delta=1+\gamma$.

Anticipating stationary profile excitations, we set $X=x-Mt$, where $M$ is the soliton speed. Poisson's equation thus leads to the pseudo-energy-balance equation
$\frac{1}{2%
}\left( d\phi /dX \right) ^{2}+\Psi (\phi,M,\kappa)=0$, where the
pseudopotential $\Psi (\phi,M,\kappa)$ is given as
\cite{sag}
\begin{equation}
\Psi(\phi,M,\kappa)   =\left(  1+ \gamma \right)  \phi+M^{2}\left[  1-\left(  1+\frac{2\phi}{M^{2}}\right)
^{1/2}\right]
 +\gamma- \gamma \left(  1-\frac{\phi}{(\kappa-\frac{3}{2}
)}\right)  ^{-\kappa+3/2}. \label{sedoveq}
\end{equation}
The lower bound of the soliton speed (sonic limit)  $M = M_s = \{(\kappa-\frac{3}{2})/[(\eta+s\delta-1)(\kappa-\frac{1}{2})]\}^{1/2}$
% $1/M_s^{2}=(\eta+s\delta-1)(\kappa-\frac{1}{2})/(\kappa-\frac{3}{2})$.
decreases with an increase in suprathermality
(i.e., decrease in $\kappa$) but increases with the value of (dust
concentration) $\delta$.
%\begin{align}
%F_{2}(M_{2}) &  =\left.  \Psi_{1}\right\vert _{\phi=-M^{2}/2}>0;\text{
%\ \ \ }U_{0}<0\nonumber\\
%&  =M_{2}^{2}\left(  1-\frac{1}{2}\left(  \eta+s\delta\right)  \right)
%+\nonumber\\
%&  +(\eta+s\delta-1)\left[  1-\left(  1+\dfrac{M_{2}^{2}}{2(\kappa-\tfrac
%{3}{2})}\right)  ^{-\kappa+3/2}\right]
%\end{align}%
% figure1[htb]
%\begin{figure}[htb]
\begin{figure}[ptb]
\centering
\includegraphics[height=1.2738in, width=5.9445in]{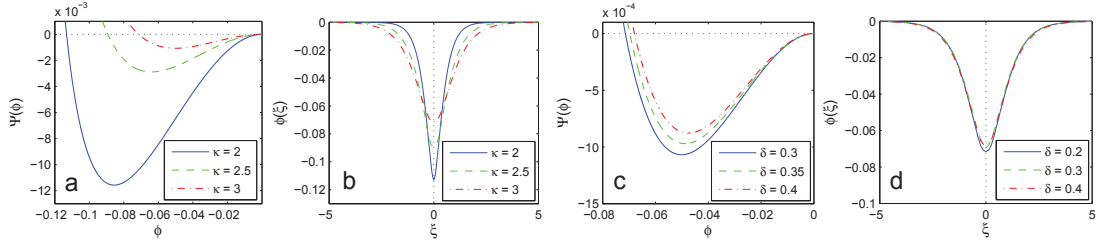}
\caption{In panels: (a) the pseudopotential $\Psi(\phi)$ and (b) the resulting pulse
profile $\phi$ are depicted, for different $\kappa$. (Here, $\delta=0.3$, $s=-1$, $\eta=5$, and $M=0.5$.)
Similarly, in panels (c) and (d), for different
values of the dust concentration $\delta$. (Here, $\kappa=3$, $s=-1$, $\eta=5$, and $M=0.5$.)} \label{figure1}
\end{figure}
As shown in Fig. \ref{figure1}, the pseudopotential becomes deeper
and wider with an increase in suprathermality and decrease in $\delta$,
for negative dusts. The pulse profile steepness also increases with
decrease in $\kappa$  and decrease in dust concentration. These effects are reversed in the presence of \emph{positive} dust.
%The combined effect of suprathermality and dust concentration plays significant role in the modification of EA soliton in dusty plasma.
We  have only observed negative polarity solitons. For $ \delta= 0$
(in the absence of dust), earlier results \cite{ad} are reproduced.

%\vskip -1cm

\section{Acknowledgments}
%{\footnotesize \textbf{Acknowledgment.}
%\textbf{Acknowledgments.}
NSS, AD and IK thank the Max-Planck Institute for Extraterrestrial
Physics for financial support. IK acknowledges support from UK EPSRC
via S\&I grant EP/D06337X/1.

%\textbf{Acknowledgments.} Organisers of the conference are warmly acknowledged for
%providing financial support to attend ICPDP6 at Garmisch-partenkirchen, Germany.

\end{document}